%% file: yoffe_ApJ_revised1.tex
\documentclass{aastex63}
\usepackage{mathrsfs}  
\usepackage{xcolor}

\accepted{November 6, 2020}
\submitjournal{ApJ}

\shorttitle{A Simplified Photodynamical Model for Planetary Mass Determination}
\shortauthors{Yoffe et al.}
\graphicspath{{./}{figures/}}

\begin{document}

\title{A Simplified Photodynamical Model for Planetary Mass Determination in Low-Eccentricity Multi-Transiting Systems}

\correspondingauthor{Gideon Yoffe}
\email{yoffe@mpia.de}

\author[0000-0002-1451-6492]{Gideon Yoffe}
\affiliation{Department of Earth and Planetary Sciences, Weizmann Institute of Science, Rehovot, 76100, Israel}
\affiliation{Now at: Max Planck Institute for Astronomy, K\"onigstuhl 17, 69117 Heidelberg, Germany}

\author[0000-0002-9152-5042]{Aviv Ofir}
\affiliation{Department of Earth and Planetary Sciences, Weizmann Institute of Science, Rehovot, 76100, Israel}

\author[0000-0001-9930-2495]{Oded Aharonson}
\affiliation{Department of Earth and Planetary Sciences, Weizmann Institute of Science, Rehovot, 76100, Israel}
\affiliation{Planetary Science Institute, Tucson, AZ, 85719, USA}



\begin{abstract}

Inferring planetary parameters from transit timing variations is challenging for small exoplanets because their transits may be so weak that determination of individual transit timing is difficult or impossible. We implement a useful combination of tools which together provide a numerically fast global photodynamical model. This is used to fit the TTV-bearing light-curve, in order to constrain the masses of transiting exoplanets in low eccentricity, multi-planet systems - and small planets in particular. We present inferred dynamical masses and orbital eccentricities in four multi-planet systems from \textit{Kepler}'s complete long-cadence data set. We test our model against Kepler-36 / KOI-277, a system with some of the most precisely determined planetary masses through TTV inversion methods, and find masses of 5.56$^{+0.41}_{-0.45}$ and 9.76$^{+0.79}_{-0.89}$ $m_\oplus$ for Kepler-36 b and c, respectively -- consistent with literature in both value and error. We then improve the mass determination of the four planets in Kepler-79 / KOI-152, where literature values were physically problematic to 12.5$^{+4.5}_{-3.6}$, 9.5$^{+2.3}_{-2.1}$, 11.3$^{+2.2}_{-2.2}$ and 6.3$^{+1.0}_{-1.0}$ $m_\oplus$ for Kepler-79 b, c, d and e, respectively. We provide new mass constraints where none existed before for two systems. These are 12.5$^{+3.2}_{-2.6}$ $m_\oplus$ for Kepler-450 c, and 3.3$^{+1.7}_{-1.0}$ and 17.4$^{+7.1}_{-3.8}$ $m_\oplus$ for Kepler-595 c (previously KOI-547.03) and b, respectively. The photodynamical code used here, called \texttt{PyDynamicaLC}, is made publicly available.
\end{abstract}



\section{Introduction}

Transit timing variations (TTVs) caused by the gravitational interactions in multi-planet systems  \citep{Holman05, Agol05} may be modeled to determine the governing system parameters, in particular masses. This is especially valuable in cases where no additional observations beyond photometry, such as radial velocity data, are available.  In practice, performing this inversion is challenging in the case of small planets and low-amplitude TTVs, although these commonly occur in nature and \textit{Kepler} data. To address this situation we combine several techniques from the literature.

\textbf{Global model}: The standard technique for the detection of TTVs follows two steps: firstly, one optimizes an empirical light-curve model using common shape parameters for all of the transit events, (in the usual \cite{MA02} framework these are  
$a/r_\star$, $b/r_\star$, and $r_p/r_\star$,
the planetary semi-major axis, impact parameter and radius, each normalized by the stellar radius)
but assigns an individual time for each transit event. Secondly, the list of best-fitting transit times is modeled with a dynamical description. There are shortcomings to this approach: the transits must be deep enough to be individually significant, they must be long enough to have more sufficient points that determine their timing, and the number of fitting parameters grows with the number of transit, greatly diminishing the sensitivity of this approach. To rectify these shortcomings \cite{Ofir18} recently introduced a more sensitive technique for the detection of TTVs: one assumes sinusoidal shape for the TTVs, and a global optimization allows determination of a single set of parameters describing the full light-curve variations. This approach is effective for small transit depths or durations (even in cases where individual transits cannot be seen), and has significantly fewer degrees of freedom - enhancing it sensitivity, discussed next.

\textbf{Parsimonious model}: The problem of inverting the observed TTVs for the component masses is non-trivial. The degeneracy between the effects of masses and orbital eccentricities has been appreciated in early works on the subject \citep{Holman05, Agol05} and a solution based on a full n-body integration is highly multi-dimensional, even more degenerate than just explained. It is thus difficult to initialize and optimize, and is computationally expensive or even prohibitive. More recently, it was shown that the pattern of the TTVs does not depend strongly on the individual eccentricities but rather only on the eccentricity-vector \textit{difference} between adjacent planetary pairs \citep{HL16, JH16, Linial8}. Fitting for the eccentricity differences allows capturing most of the TTV variability using two fewer degrees of freedom. 

We use \texttt{TTVFaster} \citep{TTVFaster} as our dynamical tool of choice. \texttt{TTVFaster} is a semi-analytic model accurate to first order in eccentricity which approximates TTVs using a series expansion. An important simplification in using this tool arises from its reliance only on \textit{average} Keplerian elements, which are well determined from previous processing steps, thus significantly decreasing the number of degrees of freedom in the model to just three per planet -- its mass and its two perpendicular eccentricity components ($e_x, e_y$, or $e\cos\omega, e\sin\omega$), where $\omega$ is the argument of periapse and $e$ is the eccentricity magnitude. Consequently, there are less  correlations among parameters, producing better determined values, and initial guesses are more easily selected. Additionally, \texttt{TTVFaster} offers a considerably faster computation time than n-body integrators (e.g. about an order of magnitude faster than \texttt{TTVFast} \citep{TTVFast}), resulting in sampler convergence timescales on the order of hours, rather than weeks \citep{Tuchow2019}. This will allow future expansion of the scope of our study to the entire sample of TTV-bearing multi-transiting \textit{Kepler} systems.

An important disadvantage of using \texttt{TTVFaster} is that its underlying approximations are inaccurate for systems with high eccentricities. Therefore, we require that our model pass a series of tests to verify the validity of our model in the relevant region of the parameter space. We also note that performing fits in flux space (the light-curve) does not permit using the linear decomposition that is available in the timing space \citep{Linial8}.

In this work we therefore build on both lessons above, and fit a model which is both global and minimal in degrees of freedom to achieve better precision and sensitivity than previous studies, and in a uniform fashion to multiple relevant \textit{Kepler} targets. We also use the recently updated stellar parameters based on Gaia DR2 \citep{Fulton2018}, which allows to better derive the planetary physical properties. The paper is divided as follows: \S\ref{processing} describes light-curve processing, in \S\ref{photodynamical_model} we describe our photodynamical model, in \S\ref{fitting} we describe our fitting procedure, in \S\ref{results} we present and discuss the results of the modeling, and in \S\ref{conclusions} we conclude.

\section{Data Processing}
\label{processing}

The sample of \textit{Kepler} systems we processed is all the multi-transiting systems of which at least one component was identified as exhibiting TTVs by \cite{Ofir18} - a total of 159 systems and 466 planets. In the first processing stage \textit{Kepler}'s photometery was iteratively detrended and the light-curve fitted using empirical TTVs (as detailed below) - to ensure the planet signals themselves do not interfere with detrending. Importantly, the goal of this stage is \textit{not} to produce a list of transit times, but only to produce the best possible detrended and normalized light-curve. The gravitational interactions among the objects are modeled in the subsequent step.

For the light-curve detrending, each of our target systems was modeled empirically in an evolved method based on \citet{Ofir2013}. All the planets in the system where fitted simultaneously using a common value $a/r_\star$, where $a$ is the semi-major axis and $r_\star$ is the stellar radius. This value was scaled between planets using Kepler's Third Law. TTVs, if exist, were accounted for by a phase shift of each event. This is a circular-orbit approximation of the TTVs. The detrending curve itself was the better of either a Savitzky-Golay filter or a cosine filter - applied to each section of continuous data independently. Each of the filters was applied several times - as many as there are known transiting planets in the system - since each transit has a unique duration. In every continuous section the transit with the longest duration appearing on that section determined the critical time-scale for filtering, and each of the filters (Savitzky-Golay or cosine) used a time-scale no shorter than three times the critical time-scale on that section.

In order to account for all TTVs we started by using the \citet{Holczer16} catalog of TTVs. In cases where measuring individual timings is difficult (shallow or short transits) individual planets were designated as not TTV-bearing. In these cases a periodic model is also likely a good approximation of the true signal (both would be difficult to detect in the first place had they exhibited large amplitude TTVs). The individual transit times were manually checked to verify that no transits were missed (e.g. \citet{Holczer16} does not report partial transits or transit close to data gaps), that no transits were unnecessarily added (some reported transit timings are at times when there is a gap in the data, for example), and in certain cases whole planets were added or removed from systems based on literature later than \citet{Holczer16}. We note that even when starting from a comprehensive catalog like \citet{Holczer16}, minimizing errors at this stage is labour intensive. For e.g., by the definition of TTVs their times are not exactly known and sometimes transits are "unexpectedly" visible in the data while the linear ephemeris suggest that the transit should fall in a data gap. Or, when transits of two similarly-sized planets that both exhibit TTVs happen to be close in time, or even overlap, correctly assigning a specific transit to a particular planet is sometimes not immediately obvious. We note this since the need for the most accurate detrending of the light-curve stems from the goal of detecting low-amplitude TTVs of shallow transits - these are inherently faint signals, and any contamination may wash away the sought-after signal altogether.

The end result of this stage is a normalized, detrended light-curve as well as a set of parameters describing the linear ephemeris for each planet: the mean period $P$, a mean reference epoch $t_{lin}$ (i.e. the constant term in the linear fit to the times of transit), $b/r_\star$, $r_p/r_\star$ and $a/r_\star$ for the innermost planet. Note, the empirically-fitted individual times of mid-transit were not used in the dynamical modeling below.

\section{Photodynamical Model}
\label{photodynamical_model}

For each multi-planet system, we compute a photodynamical model in three steps: first, we simulate a TTV signal with \texttt{TTVFaster}. The required input for each planet $i$ are the average Keplerian elements, in addition to the planetary masses and the first time of mid-transit \textbf{for a linearly approximated orbit} ($P_i$, $i_i$, $e_{x,i}$, $e_{y,i}$, $t_{{\rm lin}, i}$, $m_i$), where $i$ refers to the index of the planet, $P_i$ is the orbital period, $i$ is the orbital inclination that can be converted from $b$ with eq. 7 in \citet{Winn2011}, $e_{x,i}$ and $e_{y,i}$ are the perpendicular eccentricity components, respectively, $t_{{\rm lin}, i}$ is the mean reference epoch and $m_i$ is the mass. Note, throughout this study
the Keplerian parameters refer to their time averages over \textit{Kepler} long-cadence observations (as opposed to their osculating values).

Excluding masses and eccentricities, these parameters are well constrained during the data processing stage (\S\ref{processing}). Thus remain $3n_{\rm pl}$ degrees of freedom for the dynamical fit. The output obtained from \texttt{TTVFaster} is a vector of times of mid-transit, from which TTVs may be deduced.

Each individual transit event is then shifted from a strict periodic regime and generate the individual-event transit light-curve Mandel-Agol model \citep{MA02}, using \texttt{PyAstronomy}'s implementation of \texttt{EXOFAST}\footnote{\url{https://www.hs.uni-hamburg.de/DE/Ins/Per/Czesla/PyA/PyA/modelSuiteDoc/forTrans.html}} \citep{Exofast}. Assuming small orbital eccentricity allows us to further simplify our model and approximate the shape of the transit curve to be that of an arc of a circular orbit (referred to as a quasi-circular approximation).  
An additional assumption throughout this study is that all the TTVs can be attributed to mutual perturbations among adjacent pairs of known transiting planets. While TTVs can be sensitive to mutual inclinations \citep{Payne2010}, \textit{Kepler} multi-planet systems are usually characterized by small values of these \citep{Xie2016}, such that this effect does not make a notable contribution to TTVs of planets near low-order MMR \citep{Payne2010}.

Finally, we perform non-linear optimization of the dynamical parameters and estimate their uncertainties using \texttt{MultiNest} \citep{Multinest}, a multi-modal Markov Chain Monte Carlo algorithm with implementation in Python \citep{Buchner14}. \texttt{MultiNest} implements nested sampling \citep{Skilling04}, a Monte Carlo method targeted at the efficient calculation of the Bayesian evidence. Using \texttt{MultiNest} allows for the exploration of regions of interest in the parameter space with a reduced risk of being trapped in local likelihood maxima, which is especially important due to the known mass-eccentricity degeneracy \citep{LithXieWu12}.

\section{Model Optimization}
\label{fitting}

For any multi-planet model, we fit $3 n_{\rm pl}$ free parameters and also use $4 n_{\rm pl}+5$ input constants. The latter are the linear ephemeris (see \S\ref{processing}) and stellar parameters - $m_\star$ and $r_\star$, two stellar limb darkening coefficients and $a$ of the innermost planet. Stellar parameters are adopted from \citet{Fulton2018}. 
 
The free parameters for each planet are its mass and its two perpendicular eccentricity vector components, $e_y$ and $e_x$ (along the line of sight, and perpendicular to it along the planetary motion direction, respectively). As mentioned, in the case of small eccentricities, TTVs depend chiefly on the difference in the eccentricity vectors of each adjacent planetary pair, rather than on the individual eccentricities. Attempting to fit for both individual eccentricities yields highly degenerate results (e.g. \citet[their figures 3,5,7,9,11,15,17,19]{JH16}). Fitting for the eccentricity difference, however, results in considerably weaker mass-eccentricity degeneracy (see correlation plots in the Appendix). Therefore, we fit the absolute orbital eccentricity ($e_{x,0}$, $e_{y, 0}$) for only the innermost planet (expecting it to remain poorly constrained), 
and for the remaining planets we fit for the eccentricity differences ($\Delta e_{x, i+1} = e_{x, i+1} - e_{x, i}$, $\Delta e_{y, i+1} = e_{y, i+1} - e_{y, i}$).
 
\subsection{Statistical Model}
\label{stat_model}
 
We evaluated the goodness of fit to the multi-planet light-curve for each simulated model with parameter set $\Theta$ using the log of Student's \textit{t}-distribution function with 4 degrees of freedom, whose log is
\begin{equation}
\log(\mathscr{L}) = -\sum_{i=1} ^n \log\left(2\sqrt{2}\sigma_i \right) - \frac{5}{2}\sum_{i=1} ^n \log\left(1 + \frac{(y_{i,\Theta} - x_i)^2}{4 \sigma_i ^2} \right),
\label{Eq_LogLike}
\end{equation}
where there are $n$ observed photometric measurements $x_i$ with respective uncertainties \textbf{$\sigma_i$}, and \textbf{$y_i$} is the simulated multi-planet light-curve associated with the model parameters \textbf{$\Theta$}.

We justify the choice of our likelihood function -- Student’s t-distribution function with 4 degrees of freedom -- by comparing the distribution of residuals for all measured and best-fit model transit times in this study with two normalized and symmetric statistical distributions, a Gaussian, and Student's \textit{t}-distribution with an optimized number of degrees of freedom. We then use the Kolmogorov-Smirnov test\footnote{\url{https://docs.scipy.org/doc/scipy-0.14.0/reference/generated/scipy.stats.kstest.html}} to determine the preferred likelihood distribution, parameterized by the smallest value of $D$ - the Kolmogorov-Smirnov statistic for a given cumulative distribution. In our analysis we find that similarly to \citet{JH16, MacDonald16}, the wings of the distribution contain more outliers than would be expected by Gaussian uncertainties, such that the Student's t-distribution scored better in the Kolmogorov-Smirnov test.

\subsection{Priors}
\label{priors}

For the planetary masses, we use linear priors spanning the range  $10^{-2} m_\oplus \leq m_p \leq 10^3 m_\oplus$.

For the eccentricities, we test \textit{three} Gaussian priors in the individual eccentricity components of the inner planet and differences in eccentricity vector components 
($e_{x,0}, e_{y,0},\Delta e_{x,i}$, $\Delta e_{y,i}$)
, with zero mean and three different variances $\sigma_e =$ 0.01, 0.02 and 0.05. These  correspond to Rayleigh distributions of the eccentricity magnitude with scale-widths of 0.01, 0.02 and 0.05, respectively. Note that while the optimized parameters are the eccentricity differences, and indeed the TTV signal depends mostly on these alone, \texttt{TTVFaster} requires the absolute eccentricities of all components.

The choice of eccentricity priors is consistent with the distribution of \textit{Kepler} multi-planet systems eccentricities reported by \citet{Moorhead11} and \citet{Xie2016}, both conducting independent analyses through measured transit durations.

\subsection{Posteriors}

Our quoted values are the median values of the posterior distribution, and the uncertainties are the $50\pm 34.1$ percentiles ranges of the posterior distribution, computed after removal of the MCMC "burn in" phase (points with relative likelihood $<10^{-3}$ times that of the best-fit). We refer to the range of uncertainties as $\sigma$, defined irrespective of the distribution. For a normal distribution, $\sigma$ corresponds to the standard deviation.

Performing three optimizations with three eccentricity priors enables us to test the sensitivity of our optimization results and their respective uncertainties to the choice of eccentricity priors (following e.g. \citet{HL14, HL16, HL17}). A result for a system in which the best-fit eccentricity values are all small across all prior widths indicates that the true eccentricity is indeed small and our result is not limited by the prior.


\subsection{Validity map}
\label{ValidityMap}
To validate our fits for systems with significant non-zero eccentricities, we compare the output of \texttt{TTVFaster} to that of \texttt{TTVFast} -- a symplectic n-body integrator with a Keplerian interpolator for the calculation of transit timings, itself verified to be in good agreement with \texttt{Mercury} \citep{Mercury}. The analysis is done using a validity map we devised, in the form of a two-dimensional grid in $\Delta e_{x}$ and $\Delta e_y$ for every planet pair, keeping the remaining parameters constant at their best-fit values (producing $n_{\rm pl} - 1$ validity grids). At each point in the validity map the squared difference between \texttt{TTVFast} and \texttt{TTVFaster} normalized by the data's error estimation are used to construct a $\chi^2$-like statistic:
\begin{equation}
    \chi_{\rm validity}^2=\sum_{i=1} ^n\left(\frac{y_{i,\rm TTVFast}-y_{i,\rm TTVFaster}}{\sigma_i}\right)^2
\end{equation}
This allows evaluating the region in parameter space in which \texttt{TTVFast} and \texttt{TTVFaster} are consistent to within the measurement error.
A difficulty in performing the analysis lies in the fact that only the average Keplerian parameters are known (with uncertainty), whereas \texttt{TTVFast} requires the instantaneous values at the beginning of integration. We therefore use the average Keplerian elements as initial conditions, from which revised average values are computed from the osculating Keplerian elements. These computed averages are then used for the \texttt{TTVFaster} comparison simulation. Agreement thus also implies that the initial conditions used by first are a good approximation of the average values used by the latter. 


\subsection{Libration width and proximity to MMR}
\label{Libration}
The approximations utilized in \texttt{TTVFaster} are based on the underlying assumptions that a system satisfies three conditions: \textbf{1}. small planet-to-star mass ratios $\left( m/m_\star \lesssim 10^{-3}\right)$. \textbf{2}. small eccentricities. \textbf{3}. adjacent planets are close to, but are not in MMR \citep{TTVFaster, Weiss2017, Tuchow2019}. 


The first requirement, of small mass ratio, is verified by examination of the resulting best-fit values, which are all smaller than $10^{-3}$ for the systems presented. 

The second requirement, of small $e$, is met in systems with inferred non-zero eccentricities by performing the validity test described above (\S\ref{ValidityMap}). 

Finally, to determine the proximity of pairs of adjacent planets to MMR, we use the analytical estimate for the libration width \cite[Ch. 8]{Murray2000}.
Specifically, the term $\delta n_{\rm{max}}$ (their Eq.~8.74), the resonance width, is the maximum libration of the mean motion of a planet in resonance, which can be computed using the fitted values and choice of the appropriate resonance. For a given inner planet and resonance, $n_{\rm{res}}$ is the exactly-resonant mean motion of the outer planet. The observed mean motion $n_{\rm{obs}}$ of that outer planet is then compared to this value and we require that ${|n_{\rm{res}}-n_{\rm{obs}}|}/{\delta n_{\rm{max}}}>1$, {\it i.e.} that the planets are out of resonance. Solutions that do not meet this criterion are discarded as they violate the underlying assumptions of \texttt{TTVFaster}. We also verify the planets are near MMR in the sense that $|n_{\rm{res}}-n_{\rm{obs}}|/\delta n_{\rm{max}}$ is not large (typically of order 10 for the systems presented).

\subsection{\texttt{PyDynamicaLC}}
\label{code}

We make available our dynamical light-curve generator code for public use. The code integrates several existing routines -- \texttt{TTVFaster}, \texttt{TTVFast} (a modified C version which is also available) and \texttt{PyAstronomy}'s implementation of \texttt{EXOFAST}, sewn together to generate a light-curve with planetary dynamics in one of three configurations:

\textbf{Quasi-Circular}: dynamics are simulated with \texttt{TTVFaster}, a single transit shape is approximated as originating from a circular orbit. Each transit event is only shifted in time by the corresponding TTV determined by the dynamical simulation.

\textbf{Eccentric}: dynamics are again simulated with \texttt{TTVFaster}, the transit shapes are constant in time but with a prescribed eccentricity ($e = e$). Each transit event is shifted in time by the corresponding TTV determined by the dynamical simulation.

\textbf{Osculating}: dynamics are simulated with \texttt{TTVFast}, from which the instantaneous osculating Keplerian elements are  extracted at every mid-transit instant. The shape and time of each transit is then determined from its corresponding instantaneous elements, with the underlying assumption that Keplerian elements  during each transit remain constant.


While all configurations were validated against synthetic light-curves generated with \texttt{Mercury} \citep{Mercury} and \texttt{EXOFAST}, the work presented in this paper is based only on the quasi-circular configuration.

We further provide a coupling of PyDynamicaLC to \texttt{MultiNest}, allowing optimization of planetary masses and eccentricities following the methodology of this study, using the first two-modes of \texttt{PyDynamicaLC} (\textit{i.e.} quasi-circular and eccentric) and analyze the resulting posterior distributions.

\section{Results}
\label{results}

\subsection{Kepler-36}
\label{Kepler36}

We analyze Kepler-36 (KOI-277), a two-planet system close to a 6:7 MMR for which exceptionally precise mass constraints were previously inferred through TTV inversion by \citet{Carter2012}: both masses were determined to high precision (to better than $\sim14\sigma$). We reproduce these results in Table~\ref{Tab_KOI277_bestfits}. We chose to analyze this system as a particularly difficult test for our analysis to recover.

The mean Keplerian parameters of the system are listed in Table~\ref{Tab_KOI277_avg}. We performed three optimization rounds with different eccentricity priors, as described in \S\ref{fitting}. The best-fit values, their uncertainties and associated likelihoods, in addition to the results of \citet{Carter2012} are listed in Table~\ref{Tab_KOI277_bestfits}. We find the system to be nearly circular. In all three priors, eccentricities remain consistent with zero to within $\sim2\sigma$.

\begin{figure}
\centering
  \includegraphics[scale=0.07]{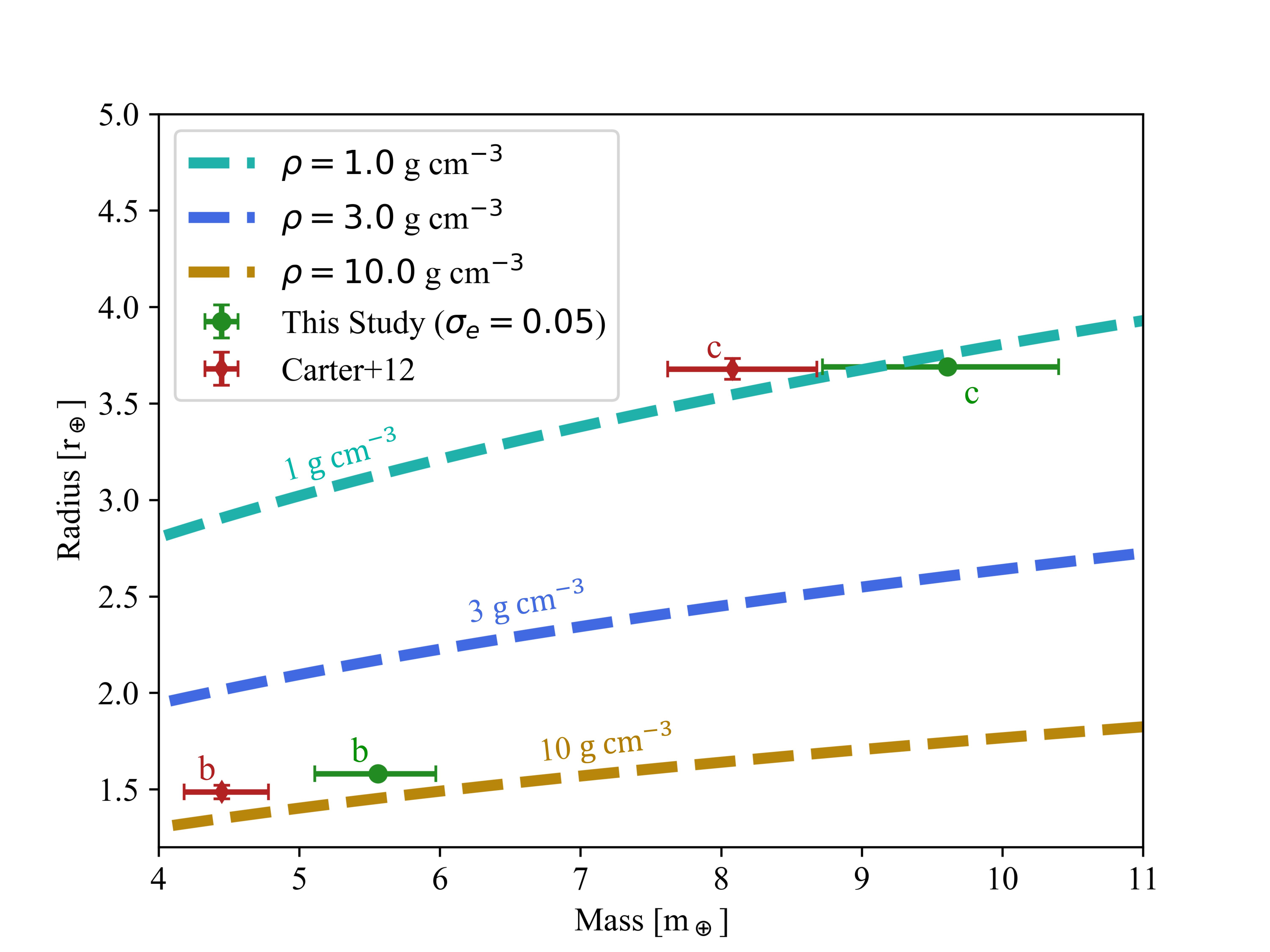}
  \caption{Kepler-36 mass-radius diagram displaying our adopted mass values (in green) and those of \citet{Carter2012} (in red), compared to three constant bulk density curves.}
  \label{Fig_KOI277_MR}
\end{figure}

We plot our adopted absolute derived planetary masses versus their radii, and those of \citet{Carter2012} in Fig.~\ref{Fig_KOI277_MR}. We find that our results are consistent with those of \citet{Carter2012} both in value and error estimates despite some differences in the underlying data sets (due to reprocessing of \textit{Kepler} data here, use of partial data by \citet{Carter2012} vs. full data set here, and inclusion of short-cadence data by \citet{Carter2012} vs. only long-cadence here) - and obviously in the data analysis techniques. We also plot the resultant posterior distributions (Fig.~\ref{Fig_KOI277_Corner}) and find that our parameterization of the eccentricity (as a single value $e_0$ and a series of differences $\Delta e_i$) is indeed nearly uncorrelated with the masses, though some correlations remain between other pairs of parameters.


\begin{table}
\small
    \centering
    \begin{tabular}{||c | c | c ||} 
    \hline
    Parameter & Kepler-36 b & Kepler-36 c \\ 
    \hline\hline
     $P$ [days] & 13.849039 $\pm$ 7.8$\cdot10^{-5}$ & 16.2318713 $\pm$ 9.4$\cdot10^{-6}$ \\
     $a/r_\star$ & 14.76 $\pm$ 0.19 & 16.41 $\pm$ 0.22 \\
     $t_{\rm lin}$ [days] & 141.7185 $\pm$ 0.0047 & 171.67304 $\pm$ 4.5$\cdot10^{-4}$  \\
     $b/r_\star$ & 0.359 $\pm$ 0.041 & 0.244 $\pm$ 0.046 \\
     $r_p$ [$r_\oplus$] & 1.582 $\pm$ 0.015 & 3.692 $\pm$ 0.010   \\
     \hline
     $m_\star$ [$m_\odot$] & \multicolumn{2}{c||}{1.034$^{+0.022}_{-0.012}$}\\
    $r_\star$ [$r_\odot$] & \multicolumn{2}{c||}{ 1.629$^{+0.020}_{-0.019}$} \\
    \hline
    \end{tabular}
    \caption{Kepler-36 geometrical model best-fit average Keplerian parameters, in addition to stellar parameters, planetary radii and their uncertainties reported in \citet{Fulton2018}.}
    \label{Tab_KOI277_avg}
\end{table}

\begin{table*}
\small
    \centering
    \begin{tabular}{||c | c | c | c | c||} 
    \hline
    $\sigma_e$ & 0.01 & 0.02 & 0.05 & Carter+12 \\ 
    \hline\hline
    m$_{b}$ [$m_\oplus$] &  5.65$^{+0.35}_{-0.48}$ & 5.61$^{+0.36}_{-0.46}$ & \textbf{5.56$^{+0.41}_{-0.45}$} & 4.45$^{+0.33}_{-0.27}$  \\
    m$_{c}$ [$m_\oplus$] & 9.76$^{+0.66}_{-0.91}$ & 9.78$^{+0.70}_{-0.88}$ & \textbf{9.61$^{+0.79}_{-0.89}$} & 8.08$^{+0.6}_{-0.46}$ \\
    e$_{x,b}$ & 0.0035$^{+0.0076}_{-0.0090}$ &  0.011$^{+0.014}_{-0.017}$ & \textbf{0.045$^{+0.023}_{-0.043}$} & $<$ 0.039  \\
    e$_{y,b}$ & -0.0031$^{+0.0092}_{-0.0082}$ &  -0.007$^{+0.018}_{-0.014}$ &  \textbf{-0.034$^{+0.045}_{-0.033}$} & $<$ 0.033 \\
    $\Delta e_{x,c}$ & 0.0117$^{+0.0069}_{-0.0082}$ & 0.019$^{+0.013}_{-0.016}$ &  \textbf{0.045$^{+0.023}_{-0.043}$} & -  \\
    $\Delta e_{y,c}$ & -0.0140$^{+0.0084}_{-0.0076}$ & -0.018$^{+0.016}_{-0.013}$ & \textbf{-0.043$^{+0.041}_{-0.030}$} & - \\
    \hline
    $-\log(\mathscr{L})$ & 496641 & 495704 & \textbf{491722} & - \\
    \hline
    \end{tabular}
    \caption{Kepler-36 derived absolute planetary masses, absolute eccentricity components of the inner planet and eccentricity component differences for the outer planets. Adopted values are in bold and results of \citet{Carter2012} are quoted in the rightmost column for comparison (see also Fig.~\ref{Fig_KOI277_MR}.)}
    \label{Tab_KOI277_bestfits}
\end{table*}

\subsection{Kepler-79}
\label{Kepler79}

We analyze Kepler-79 (KOI-152), a four-planet system close to the 2:1, 2:1 and 3:2 MMRs, respectively. The average Keplerian parameters for the system are listed in Table~\ref{Tab_KOI152_avg}. We performed three optimization rounds as described in \S\ref{priors}. The adopted values, their uncertainties and associated likelihoods, in addition to the adopted values of \citet{JH13} are listed in in Table~\ref{Tab_KOI152_bestfits}.

\begin{figure}
\centering
  \includegraphics[scale=0.07]{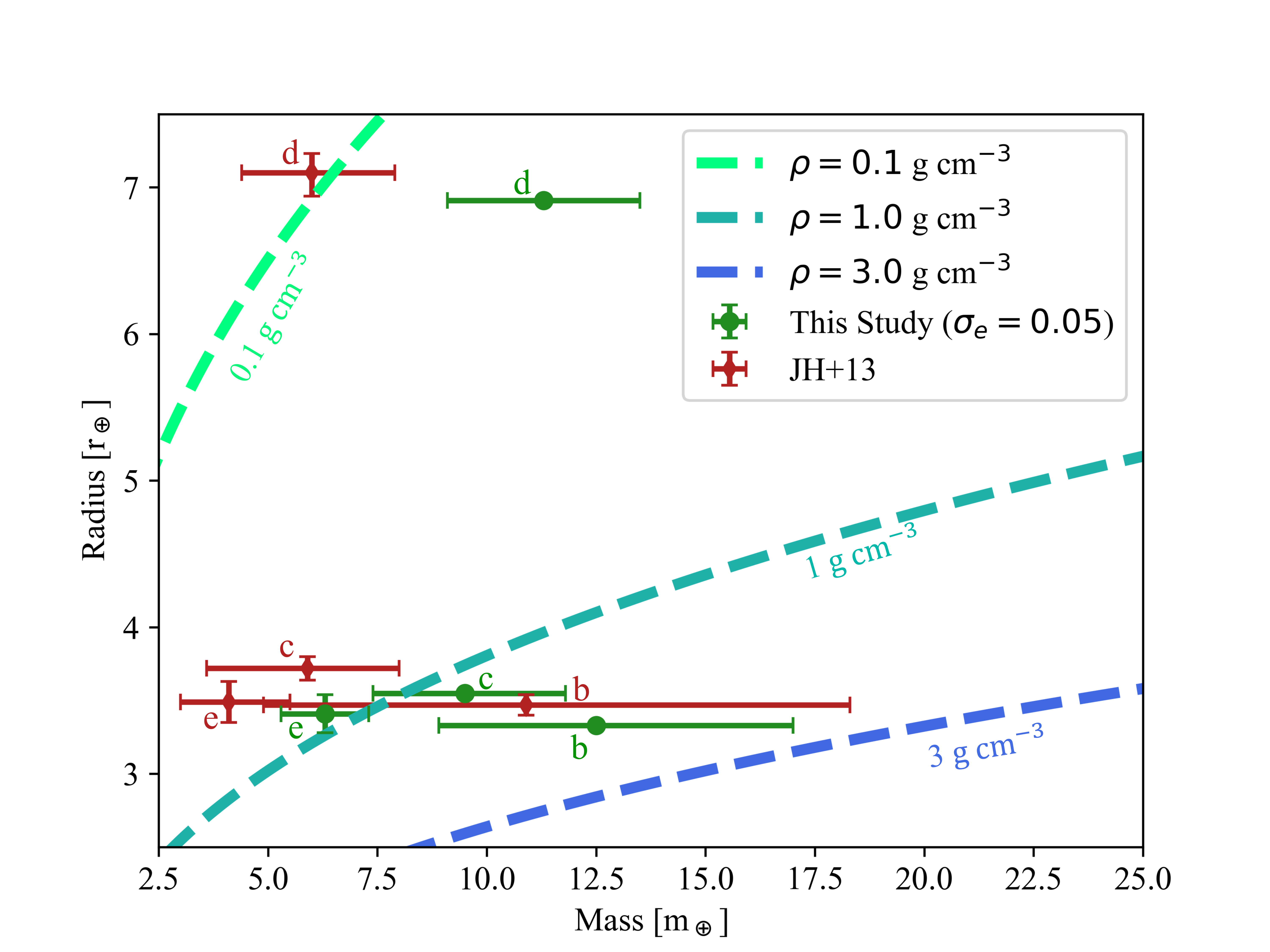}
  \caption{Kepler-79 mass-radius diagram displaying our adopted mass values (in green) and those of \citet{JH13} (in red), compared to three constant bulk density curves.}
  \label{Fig_KOI152_MR}
\end{figure}

For this system, mass constraints were previously inferred through TTV inversion by \citet{JH13}. In their analysis, \citet{JH13} find that the four planets have low bulk densities, in particular Kepler-79 d. In this case, our approach allows us to decrease the mass uncertainty of Kepler-79 b by a factor of $\sim$2 for the innermost and smallest Kepler-79 b, and increase the mass of Kepler-79 d, moving it to a more physically plausible location on the mass-radius diagram (Fig.~\ref{Fig_KOI152_MR}). We also plot the resultant posterior distributions (Fig.~\ref{Fig_KOI152_Corner}) and find some, but not strong, mass-eccentricity correlations.

Similarly to Kepler-36 (\S\ref{Kepler36}), we find all planets in this system have eccentricities that are consistent with zero. For 12 degrees of freedom, the variations in $- \log(\mathscr{L})$ of the different optimization runs are statistically insignificant. We choose the $\sigma_e = 0.05$ optimization run as our adopted values based on its lowest $- \log(\mathscr{L})$ value, though as mentioned, the improvement is small. 

\begin{table*}
\small
    \centering
    \begin{tabular}{||c | c | c | c | c||} 
    \hline
    Parameter & Kepler-79 b & Kepler-79 c & Kepler-79 d & Kepler-79 e \\ 
    \hline\hline
     $P$ [days] & 13.484542 $\pm$ 2.1$\cdot10^{-5}$ & 27.402300 $\pm$ 4.2$\cdot10^{-5}$ & 52.090767 $\pm$ 3.2$\cdot10^{-5}$ & 81.063633 $\pm$ 4.4$\cdot10^{-4}$  \\
     $a/r_\star$ & 19.37 $\pm$ 0.25 & 31.08 $\pm$ 0.40 & 47.69 $\pm$ 0.61 & 64.05 $\pm$ 0.82  \\
     $t_{\rm lin}$ [days] & 136.6209 $\pm$ 0.0012 & 133.6276 $\pm$ 0.0013 & 158.74645 $\pm$ 5.4$\cdot10^{-4}$ & 139.5377 $\pm$ 0.0045  \\
     $b/r_\star$ & 0.360 $\pm$ 0.035 & 0.057 $\pm$ 0.077 & 0.043 $\pm$ 0.076 & 0.934 $\pm$ 0.016  \\
     $r_p$ [$r_\oplus$] & 3.338 $\pm$ 0.028 & 3.553 $\pm$ 0.027 & 6.912 $\pm$ 0.014 & 3.414 $\pm$ 0.129  \\
    \hline
    $m_\star$ [$m_\odot$] & \multicolumn{4}{c||}{1.244$^{+0.027}_{-0.042}$}  \\
    $r_\star$ [$r_\odot$] & \multicolumn{4}{c||}{1.283$^{+0.029}_{-0.015}$} \\
    \hline
    \end{tabular}
    \caption{Kepler-79 geometrical model best-fit average Keplerian parameters, in addition to stellar parameters, planetary radii and their uncertainties reported in \citet{Fulton2018}.}
    \label{Tab_KOI152_avg}
\end{table*}

\begin{table*}
\small
    \centering
    \begin{tabular}{||c | c | c | c | c||} 
    \hline
    $\sigma_e$ & 0.01 & 0.02 & 0.05 & JH+13 \\ 
    \hline\hline
    m$_{b}$ [$m_\oplus$] &  17.8$^{+5.4}_{-3.8}$ & 15.0$^{+4.5}_{-3.6}$ & \textbf{12.5$^{+4.5}_{-3.6}$} & 10.9$^{+7.4}_{-6.0}$  \\
    m$_{c}$ [$m_\oplus$] & 12.4$^{+2.1}_{-1.8}$ & 11.0$^{+2.1}_{-1.9}$ & \textbf{9.5$^{+2.3}_{-2.1}$} & 5.9$^{+1.9}_{-2.3}$ \\
    m$_{d}$ [$m_\oplus$] & 13.2$^{+2.1}_{-1.9}$ & 12.3$^{+1.9}_{-1.9}$ & \textbf{11.3$^{+2.2}_{-2.2}$} & 6.0$^{+2.1}_{-1.6}$ \\
    m$_{e}$ [$m_\oplus$] & 6.85$^{+0.91}_{-0.88}$ & 6.53$^{+0.86}_{-0.87}$ & \textbf{6.3$^{+1.0}_{-1.0}$} & 4.1$^{+1.4}_{-1.1}$ \\
    e$_{x,b}$ & -0.0125$^{+0.0026}_{-0.0030}$ & -0.0148$^{+0.0033}_{-0.0039}$ & \textbf{-0.0173$^{+0.0046}_{-0.0061}$} & 0.032$^{+0.036} _{-0.032}$  \\
    e$_{y,b}$ & 0.0014$^{+0.0020}_{-0.0019}$ & 0.0007$^{+0.0026}_{-0.0025}$ & \textbf{-0.0007$^{+0.0038}_{-0.0038}$} & 0.032$^{+0.059} _{-0.029}$ \\
    $\Delta e_{x,c}$ & -0.0145$^{+0.0040}_{-0.0044}$ & -0.0171$^{+0.0051}_{-0.0056}$ & \textbf{-0.0195$^{+0.0072}_{-0.0081}$} & -  \\
    $\Delta e_{y,c}$ &  -0.0065$^{+0.0051}_{-0.0053}$ & -0.0123$^{+0.0072}_{-0.0079}$ & \textbf{-0.021$^{+0.011}_{-0.015}$} & - \\
    $\Delta e_{x,d}$ & -0.0108$^{+0.0075}_{-0.0077}$ &  -0.014$^{+0.012}_{-0.012}$ & \textbf{-0.015$^{+0.021}_{-0.021}$} & -  \\
    $\Delta e_{y,d}$ & 0.0028$^{+0.0076}_{-0.0074}$ &  -0.004$^{+0.013}_{-0.012}$ & \textbf{-0.018$^{+0.021}_{-0.023}$} & -\\
    $\Delta e_{x,e}$ & -0.0129$^{+0.0062}_{-0.0066}$ &  -0.015$^{+0.010}_{-0.010}$ &\textbf{ -0.017$^{+0.017}_{-0.017}$} & -  \\
    $\Delta e_{y,e}$ & -0.0067$^{+0.0063}_{-0.0063}$ & -0.014$^{+0.011}_{-0.011}$ & \textbf{-0.026$^{+0.018}_{-0.020}$} & - \\
    \hline
    $-\log(\mathscr{L})$ & 443838 & 443819 & \textbf{443806} & - \\
    \hline
    \end{tabular}
    \caption{Kepler-79 derived absolute planetary masses, absolute eccentricity components of the inner planet and eccentricity component differences for the outer planets. Adopted values are boldfaced and results of \citet{JH13} are quoted in the rightmost column.}
    \label{Tab_KOI152_bestfits}
\end{table*}

\subsection{Kepler-450}

We analyze Kepler-450 (KOI-279), a three-planet system close to the 2:1 MMR for both adjacent pairs. To our knowledge, there exist no previous mass or eccentricity constraints in the literature for any of the planets in the system. The average Keplerian parameters for the system are listed in Table~\ref{Tab_KOI279_avg}. We again performed three optimization runs. The fitting results, their uncertainties and associated likelihoods are listed in Table~\ref{Tab_KOI279_bestfits}. 

\begin{figure}
\centering
  \includegraphics[scale=0.07]{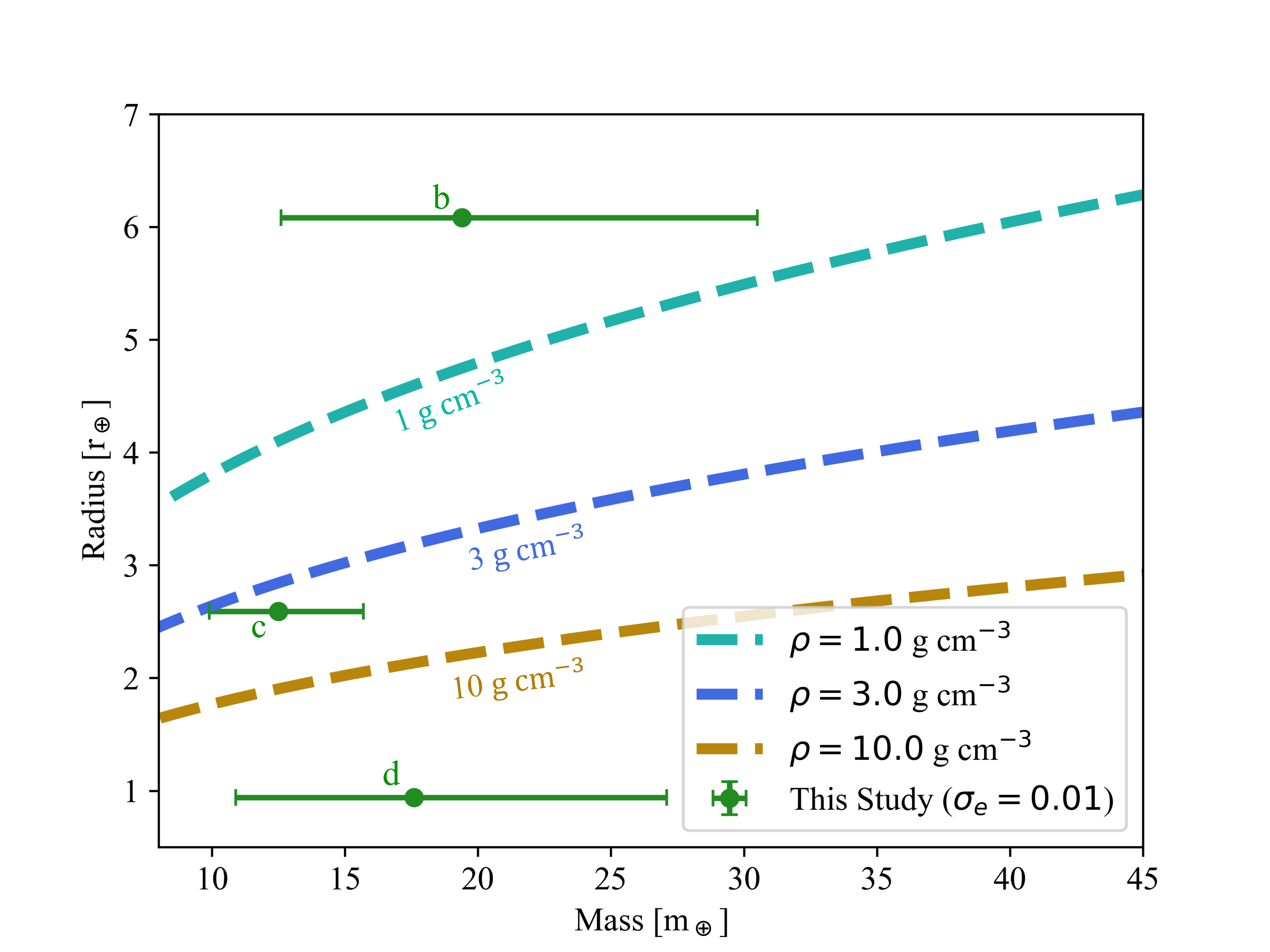}
  \caption{Kepler-450 mass-radius diagram displaying our adopted mass values compared to three constant bulk density curves. The best-fit masses of Kepler-450 b and Kepler-450 d are anomalous for their sizes, but are poorly constrained.}
  \label{Fig_KOI279_MR}
\end{figure}

\begin{figure}
\centering
  \includegraphics[scale=0.11]{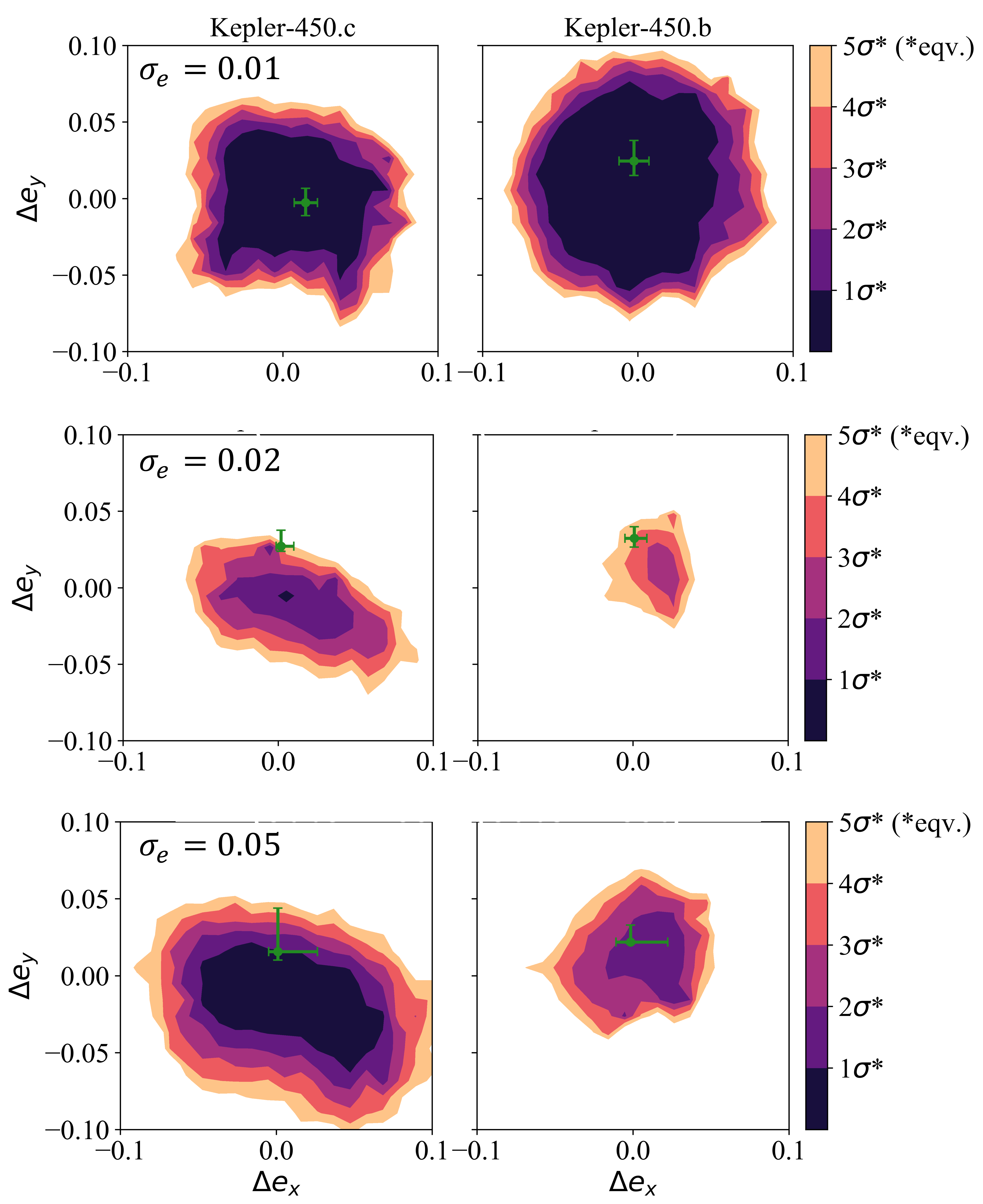}
  \caption{Validity maps for Kepler-450. Best-fit solutions and their uncertainties are marked in green. Each row refers to an optimization run with a different eccentricity prior scale-width ($\sigma_e$). Here, the best-fit parameters are kept constant (Table~\ref{Tab_KOI279_bestfits}) except the $\Delta e_x, \Delta e_y$ values of each of the outer planets scanned in each column. Contours are of the $\chi^2_{\rm validity}$  (\S\ref{ValidityMap}) measuring the difference between \texttt{TTVFast} and \texttt{TTVFaster} generated light-curves. The values are converted to the equivalent number of standard deviations assuming a normal distribution with $3n_{\rm pl}$ degrees of freedom. Best-fit values from each optimization run and their uncertainties are marked with a green cross.}
  \label{Fig_Validity_KOI279_1}
\end{figure}

In the case of Kepler-450, the results from the optimization runs assuming the two wider eccentricity prior distributions indicate small but significantly non-zero eccentricities. Therefore, we perform our validity map analysis, varying  $\Delta e_x, \Delta e_y$ near their best-fit values. This analysis shows that the fit lies, in fact, in a region showing 2-5$\sigma$ disagreement between the \texttt{TTVFast} and \texttt{TTVFaster} models (see Fig.~\ref{Fig_Validity_KOI279_1}). This suggests that in these regions of parameter space \texttt{TTVFaster} model may not be reliable . The $\sigma_e = 0.01$ optimization run does, however, lie well within a favorable region, with $< 1\sigma$ disagreement between the two calculations (Fig.~\ref{Fig_Validity_KOI279_1}). Consequently, we choose this solution as our preferred values for this system. We plot our adopted planetary masses versus their radii in Fig.~\ref{Fig_KOI279_MR}. Note, that while the masses Kepler-450 d and Kepler 450 b appear to be physically implausible -- too dense in the case of the first and too light in the case of the latter, both have large error bars, and hence we do not attach significance to their mass constraints. We also plot the resultant posterior distributions (Fig.~\ref{Fig_KOI279_Corner}) and here too find little to no mass-eccentricity correlation using our parameterization.

\begin{table}
\small
    \centering
    \begin{tabular}{||c | c | c | c||} 
    \hline
    Parameter & Kepler-450 d & Kepler-450 c & Kepler-450 b\\ 
    \hline\hline
     $P$ [days] & 7.5144279 $\pm$ 5.5$\cdot10^{-6}$ & 15.4131395 $\pm$ 2.9$\cdot10^{-6}$ & 28.4548844 $\pm$ 1.0$\cdot10^{-6}$ \\
     $a/r_\star$ & 10.859 $\pm$ 0.012 & 17.531 $\pm$ 0.020 & 26.383 $\pm$ 0.031 \\
     $t_{\rm lin}$ [days] & 136.17796 $\pm$ 5.0$\cdot10^{-4}$ & 136.94162 $\pm$ 1.1$\cdot10^{-4}$ & 176.705831 $\pm$ 2.1$\cdot10^{-5}$  \\
     $b/r_\star$ & 0.5214 $\pm$ 0.0033 & 0.3007 $\pm$ 0.0027 & 0.3038 $\pm$ 0.0032 \\
     $r_p$ [$r_\oplus$] & 0.9408 $\pm$ 0.0018 & 2.5958 $\pm$ 0.0017 & 6.0834 $\pm$ 0.0022  \\
     \hline
    $m_\star$ [$m_\odot$] & \multicolumn{3}{c||}{1.334$^{+0.023}_{-0.022}$} \\
    $r_\star$ [$r_\odot$] & \multicolumn{3}{c||}{1.6$^{+0.028}_{-0.008}$} \\
    \hline
    \end{tabular}
    \caption{Kepler-450 geometrical model best-fit average Keplerian parameters, in addition to stellar parameters, planetary radii and their uncertainties reported in \citet{Fulton2018}.}
    \label{Tab_KOI279_avg}
\end{table}

\begin{table*}
\small
    \centering
    \begin{tabular}{||c | c | c | c||} 
    \hline
    $\sigma_e$ & 0.01 & 0.02 & 0.05 \\ 
    \hline\hline
    m$_{d}$ [$m_\oplus$] &  \textbf{17.6$^{+9.5}_{-6.7}$} & 11.4$^{+6.8}_{-5.1}$ & 7.5$^{+6.8}_{-3.5}$  \\
    m$_{c}$ [$m_\oplus$] & \textbf{12.5$^{+3.2}_{-2.6}$} & 55$^{+30}_{-31}$ & 33$^{+35}_{-24}$ \\
    m$_{b}$ [$m_\oplus$] & \textbf{19.4$^{+11.1}_{-6.8}$} &  30$^{+16}_{-13}$ & 25$^{+14}_{-11}$ \\
    e$_{x,d}$ & \textbf{0.0079$^{+0.0049}_{-0.0040}$} &  0.0037$^{+0.0054}_{-0.0031}$ & 0.0058$^{+0.0158}_{-0.0050}$  \\
    e$_{y,d}$ & \textbf{0.0128$^{+0.0051}_{-0.0061}$} &  0.0303$^{+0.0084}_{-0.0069}$ &  0.039$^{+0.022}_{-0.014}$ \\
    $\Delta e_{x,c}$ & \textbf{0.0194$^{+0.0077}_{-0.0073}$} & 0.0059$^{+0.0083}_{-0.0031}$ &  0.0094$^{+0.0254}_{-0.0059}$  \\
    $\Delta e_{y,c}$ & \textbf{0.0108$^{+0.0085}_{-0.0094}$} &  0.0451$^{+0.0034}_{-0.0104}$ & 0.0422$^{+0.0055}_{-0.0283}$ \\
    $\Delta e_{x,b}$ & \textbf{0.0071$^{+0.0097}_{-0.0096}$} & 0.0046$^{+0.0082}_{-0.0059}$ &  0.0076$^{+0.0239}_{-0.0093}$  \\
    $\Delta e_{y,b}$ & \textbf{0.0245$^{+0.0095}_{-0.0134}$} & 0.0541$^{+0.0057}_{-0.0075}$ & 0.056$^{+0.011}_{-0.011}$  \\
    \hline
    $-\log(\mathscr{L})$ & \textbf{500095} & 500099 & 500098 \\
    \hline
    \end{tabular}
    \caption{Kepler-450 derived absolute planetary masses, absolute eccentricity components of the inner planet and eccentricity component differences for the outer planets. Adopted values are boldfaced.}
    \label{Tab_KOI279_bestfits}
\end{table*}

\subsection{Kepler-595}

We analyze Kepler-595 (KOI-547), which exhibits three sets of transit signals with pairs close to the 5:3 and 2:1 MMRs. To our knowledge, there exist no previous mass or eccentricity constraints for any of the planets in the system, and while the outer planet is statistically validated \citep{Morton_2016} the two inner planets are still labeled as candidates. The average Keplerian parameters for the system are listed in Table~\ref{Tab_KOI547_avg}. The best-fit values of the three optimization runs, their uncertainties and associated likelihoods are listed in Table~\ref{Tab_KOI547_bestfits}. 

\begin{figure}
\centering
  \includegraphics[scale=0.11]{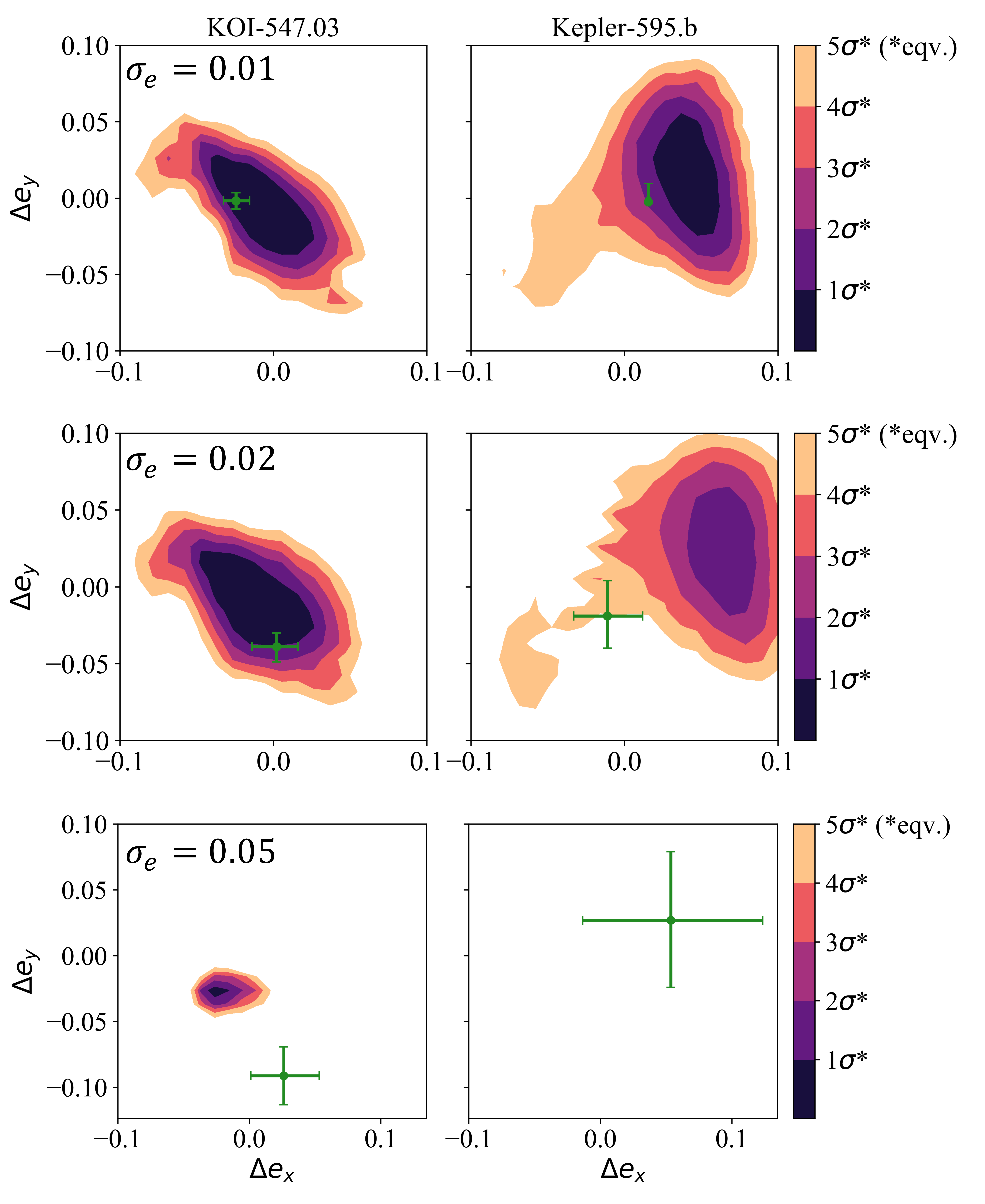}
  \caption{Validity maps for Kepler-595, similar to Fig.~\ref{Fig_Validity_KOI279_1}}
  \label{Fig_Validity_KOI547_1}
\end{figure}

Kepler-595 too requires the validity map test for results with significant non-zero eccentricities. Figure \ref{Fig_Validity_KOI547_1} shows that the two lowest $-\log(\mathscr{L})$ solutions, for $\sigma_e = 0.02, 0.05$, lie in regions of the parameter-space where \texttt{TTVFast} and \texttt{TTVFaster} models are not in agreement. The $\sigma_e = 0.01$ optimization run, however, lies in a more favorable region. We therefore choose the values from this run as our adopted values for this system. Similarly to the case of Kepler-450 d, the best-fit mass of KOI-547.02 appears to be physically implausible, but it too has large error bars (consistent with zero at 2$\sigma$).

\begin{figure}
\centering
\includegraphics[scale=0.07]{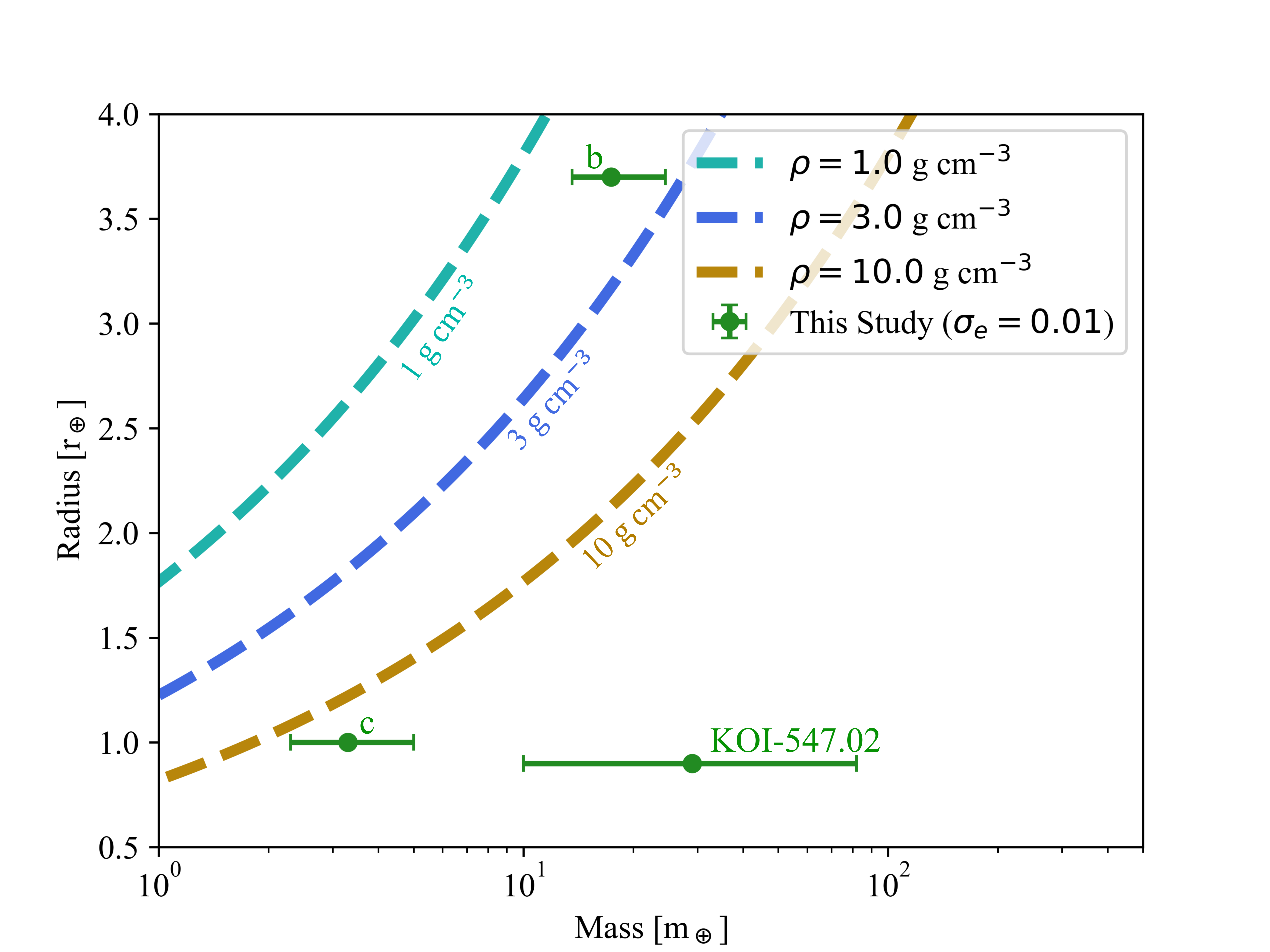}
\caption{Kepler-595 mass-radius diagram displaying our adopted mass values compared to three constant bulk density curves. Error-bars denote 1$\sigma$ uncertainty in $m$ and $r_p$. Note the mass of KOI-547.02 is poorly constrained.}
\label{Fig_KOI547_MR}
\end{figure}

We plot our preferred absolute derived planetary masses versus their radii in Fig.~\ref{Fig_KOI547_MR}, and the resultant posterior distributions (Fig.~\ref{Fig_KOI547_Corner}). Here the mass-eccentricity correlation is evident even in our eccentricity parameterization - possibly reflecting the dearth of true mass information.

\begin{table*}
\small
    \centering
    \begin{tabular}{||c | c | c | c||} 
    \hline
    Parameter & KOI-547.02 & KOI-547.03 & Kepler-595 b \\ 
    \hline\hline
     $P$ [days] & 7.3467925 $\pm$ 5.9$\cdot10^{-6}$ & 12.38602 $\pm$ 5.0$\cdot10^{-6}$ & 25.3029092 $\pm$ 6.0$\cdot10^{-6}$ \\
     $a/r_\star$ & 19.613 $\pm$ 0.035 & 27.783 $\pm$ 0.049 & 44.730 $\pm$ 0.080 \\
     $t_{\rm lin}$ [days] & 134.5069 $\pm$ 0.0052 & 132.8192 $\pm$ 0.0037 & 188.06096 $\pm$ 1.0$\cdot10^{-4}$  \\
     $b/r_\star$ & 0.593 $\pm$ 0.052 & 0.710 $\pm$ 0.063 & 0.206 $\pm$ 0.051 \\
     $r_p$ [$r_\oplus$] & 0.952 $\pm$ 0.015 & 1.009 $\pm$ 0.024 & 3.708 $\pm$ 0.010  \\
     \hline
    $m_\star$ [$m_\odot$] & \multicolumn{3}{c||}{0.93 $\pm$ 0.059} \\
    $r_\star$ [$r_\odot$] & \multicolumn{3}{c||}{0.821 $\pm$ 0.242} \\
    \hline
    \end{tabular}
    \caption{Kepler-595 geometrical model best-fit average Keplerian parameters, in addition to stellar parameters, planetary radii and their uncertainties are from \textit{NexSci}.}
    \label{Tab_KOI547_avg}
\end{table*}

\begin{table}
\small
    \centering
    \begin{tabular}{||c | c | c | c||} 
    \hline
    $\sigma_e$ & 0.01 & 0.02 & 0.05 \\ 
    \hline\hline
    m$_{547.02}$ [$m_\oplus$] &  \textbf{29$^{+53}_{-19}$} & 35$^{+75}_{-25}$ & 126$^{+42}_{-41}$  \\
    m$_{547.03}$ [$m_\oplus$] &  \textbf{3.3$^{+1.7}_{-1.0}$} & 2.34$^{+1.57}_{-0.85}$ & 2.59$^{+0.94}_{-0.81}$ \\
    m$_{b}$ [$m_\oplus$] & \textbf{17.4$^{+7.1}_{-3.8}$} & 14.9$^{+9.4}_{-5.0}$ & 23.1$^{+8.3}_{-7.8}$ \\
    e$_{x,547.02}$ & \textbf{-0.0108$^{+0.0074}_{-0.0063}$} &  -0.021$^{+0.014}_{-0.012}$ &  -0.060$^{+0.037}_{-0.034}$  \\
    e$_{y,547.02}$ & \textbf{0.0055$^{+0.0070}_{-0.0080}$} &  0.017$^{+0.014}_{-0.015}$ &  0.094$^{+0.031}_{-0.037}$ \\
    $\Delta e_{x,547.03}$ & \textbf{-0.0219$^{+0.0088}_{-0.0082}$} & -0.032$^{+0.014}_{-0.016}$ &  -0.042$^{+0.027}_{-0.025}$  \\
    $\Delta e_{y,547.03}$ & \textbf{0.0027$^{+0.0054}_{-0.0052}$} & 0.0078$^{+0.0098}_{-0.0090}$ & 0.028$^{+0.022}_{-0.022}$ \\
    $\Delta e_{x,b}$ & \textbf{-0.007$^{+0.011}_{-0.011}$} &  -0.015$^{+0.023}_{-0.022}$ &  -0.043$^{+0.070}_{-0.067}$  \\
    $\Delta e_{y,b}$ &  \textbf{0.007$^{+0.012}_{-0.012}$} & 0.018$^{+0.021}_{-0.023}$ &  0.079$^{+0.051}_{-0.052}$  \\
    \hline
    $-\log(\mathscr{L})$ & \textbf{407075} & 407072 & 406992 \\
    \hline
    \end{tabular}
    \caption{Kepler-595 derived absolute planetary masses, absolute eccentricity components of the inner planet and eccentricity component differences for the outer planets. Adopted values are boldfaced.}
    \label{Tab_KOI547_bestfits}
\end{table}

\section{Conclusions}
\label{conclusions}



We implemented a simplified photodynamical model to infer planetary masses and eccentricities in multi-planet systems. The approach couples dynamical output simulated with \texttt{TTVFaster} \citep{TTVFaster}, a semi-analytic TTV model, with a \citet{MA02} analytical light-curve model. Optimization of planetary parameters was done by \texttt{MultiNest} \citep{Multinest}, a multi-modal Bayesian inference algorithm, applied using a unique eccentricity parametrization: the eccentricity \textit{differences} between adjacent planets as our primary degrees of freedom, which enables the optimization to be significantly less prone to correlations. The median CPU-time that was required for the systems analyzed is $\sim$6.5 hr per system (single-threaded). The dynamical optimization is thus not CPU-intensive for these few-planet systems. Overall, the reduced number of degrees of freedom in \texttt{TTVFaster}, the combination of the choice of jump parameters and the fitting the light-curve and not the transit timings improved the mass sensitivity here, which allowed the analysis of a sample of \textit{Kepler} multi-planet systems, and in particular planets that were thus far too small to have well-constrained transit timings. Finally, we quality-check our best-fit values by ensuring that they are at a region of parameter space where \texttt{TTVFaster} is consistent with \texttt{TTVFast}, and that the suggested \texttt{TTVFaster} solution is close to, but not in, resonance.


We were able to provide significant planetary mass (and eccentricity) constraints in several planetary systems: we decreased the uncertainty of Kepler-79 b by a factor of $\sim$2, and our derived mass values of Kepler-79 d suggests more physically plausible bulk density. We also provide the first mass constraints to Kepler-595 b, Kepler-450 c and KOI-547.03 - a Neptune-sized , sub-Neptune and an Earth-sized planets respectively. The mass of Kepler-450 b is marginally detected to $2.9\sigma$. The simplified photodynamical code used in this study is made publicly available \footnote{\url{https://github.com/AstroGidi/PyDynamicaLC}}.

Looking forward, we plan to use this novel and now validated technique to analyze all multi-transiting systems from the \textit{Kepler} and \textit{TESS} data sets with known TTVs.

\section*{Acknowledgements}

We wish to thank Dr. Trifon Trifonov and Yair Judkovsky for helpful discussions, and Prof. Eric Agol for advice on \texttt{TTVFaster} and \texttt{TTVFast}. This study was supported by the Helen Kimmel Center for Planetary Sciences and the Minerva Center for Life Under Extreme Planetary Conditions \#13599 at the Weizmann Institute of Science.

\input{abbrev}

\bibliographystyle{mnras}
\bibliography{myref}

\appendix

\section{Correlation Plots}

In the following section we provide a plot for each system considered, in which pairwise correlations are displayed.

\begin{figure*}
\centering
  \includegraphics[scale=0.4]{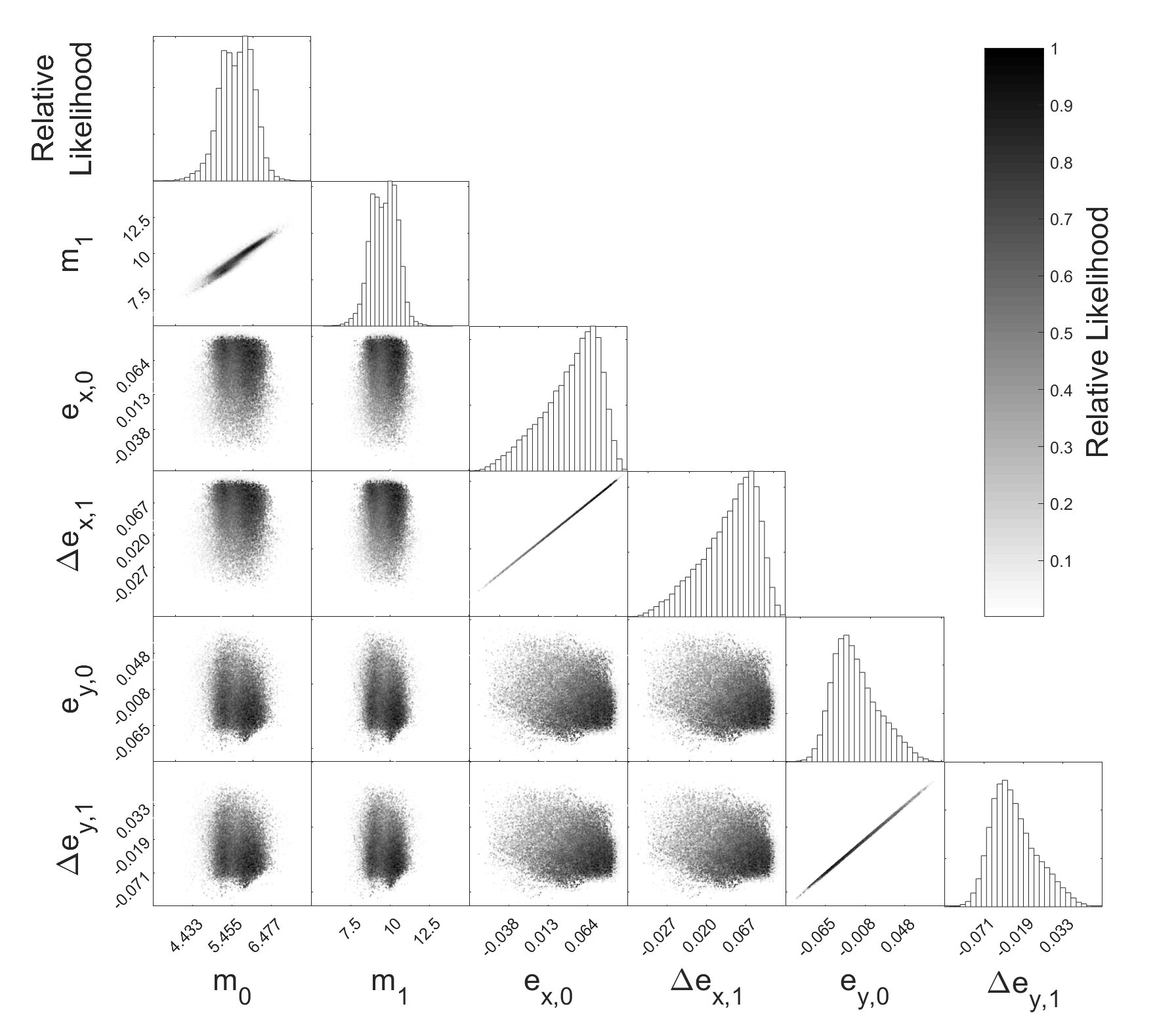}
  \caption{Kepler-36 correlation plot. The color bar is the relative likelihood of each point, relative to the best-fit point in the posterior sample (Relative Likelihood $\propto\exp(-\Delta \log\mathscr{L}$)).}
  \label{Fig_KOI277_Corner}
\end{figure*}

\begin{figure*}
\centering
  \includegraphics[scale=0.4]{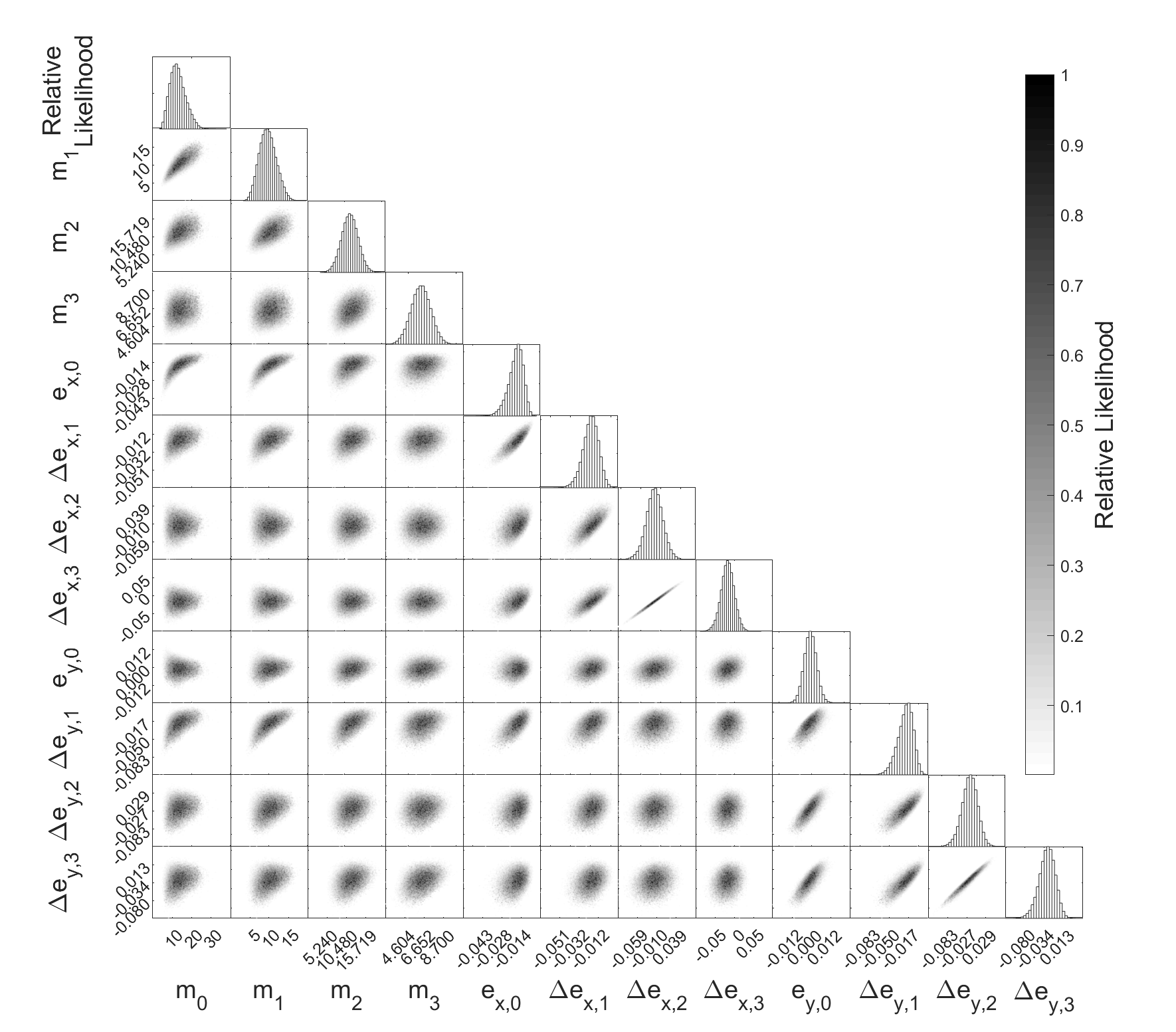}
  \caption{Kepler-79 correlation plot, similarly to Fig.~\ref{Fig_KOI277_Corner}.}
  \label{Fig_KOI152_Corner}
\end{figure*}

\begin{figure*}
\centering
  \includegraphics[scale=0.4]{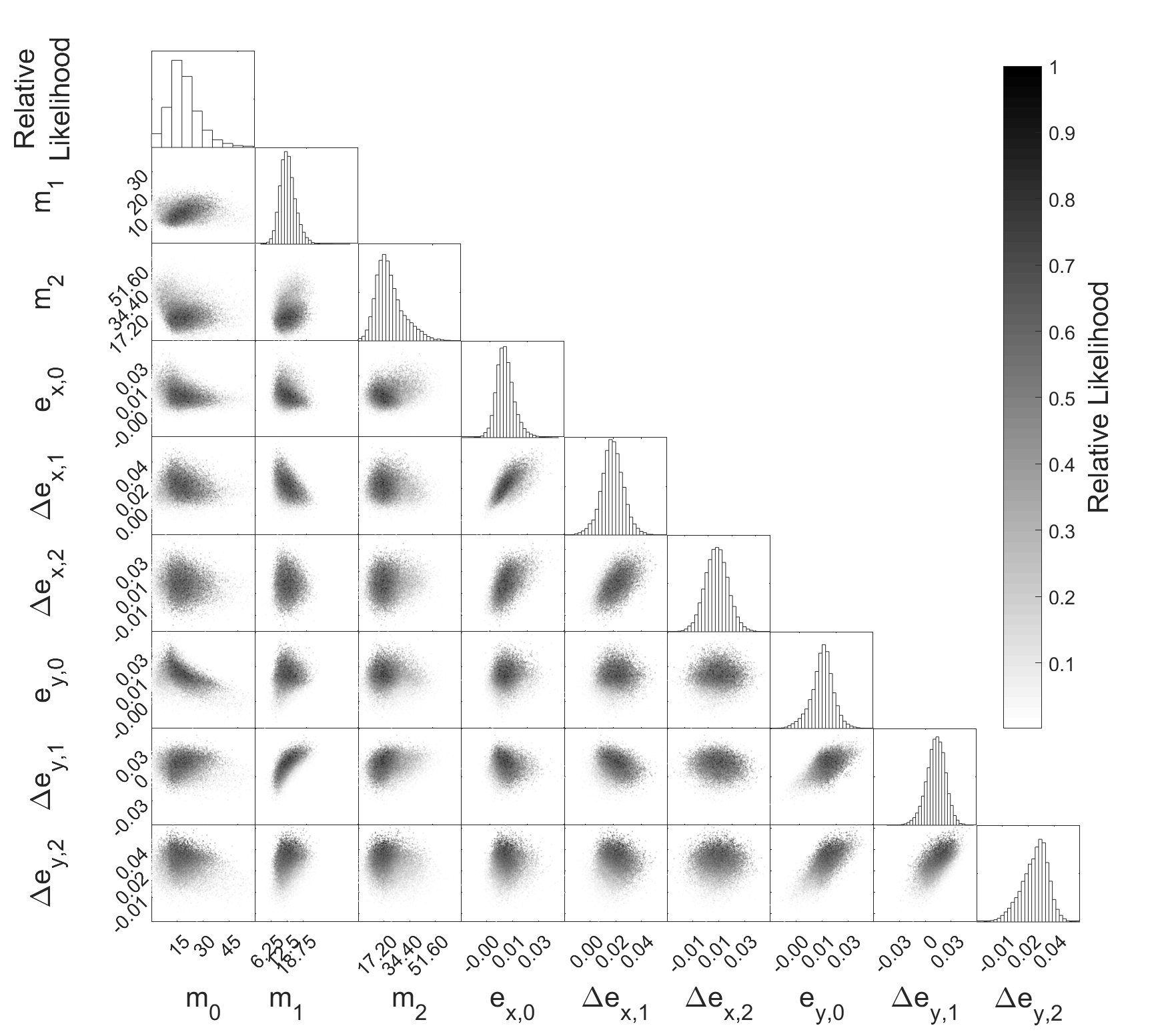}
  \caption{Kepler-450 correlation plot, similarly to Fig.~\ref{Fig_KOI277_Corner}.}
  \label{Fig_KOI279_Corner}
\end{figure*}

\begin{figure*}
\centering
  \includegraphics[scale=0.4]{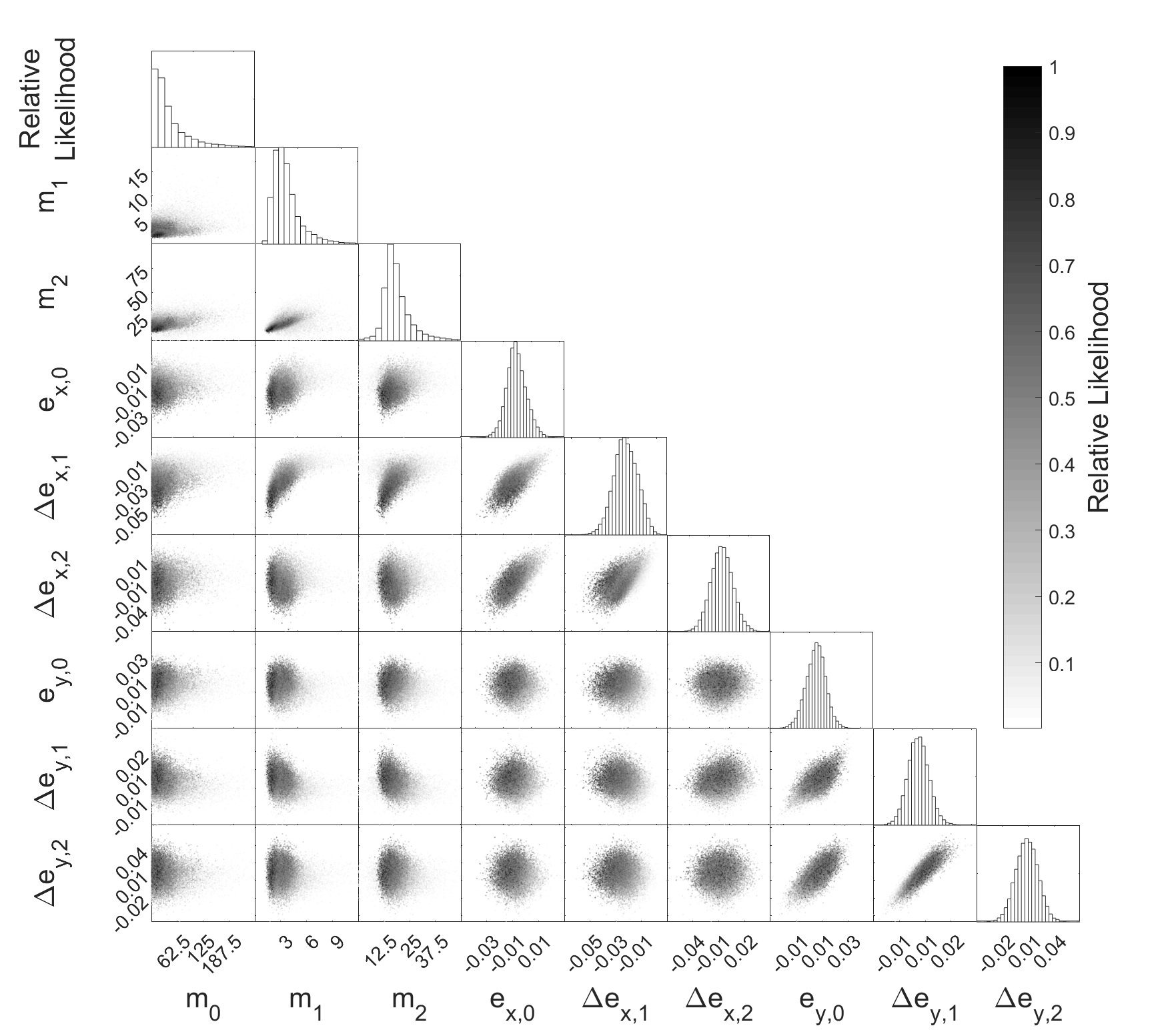}
  \caption{Kepler-595 correlation plot, similarly to Fig.~\ref{Fig_KOI277_Corner}.}
  \label{Fig_KOI547_Corner}
\end{figure*}

\label{lastpage}
\end{document}

%% file: abbrev.tex
\def\mnras{Month. Not. Royal Acad. Soc.}
\def\apj{Astrophys. J.}
\def\aap{Astro. Astrophys.}
\def\apjl{Astrophys. J. Let.}
\def\physrep{Physics Reports}
\def\aj{Astronomical Journal}
\def\apjs{Astrophys. J., Supp.}
\def\nat{Nature}
\def\icarus{Icarus}
\def\solphys{Solar Physics}
\def\araa{Ann. Rev. of Astron and Astrophys}